\documentclass[12pt,showpacs,amssymb,superscriptaddress]{revtex4}
\usepackage{mathrsfs}

\usepackage{amssymb,amsmath}
\usepackage{amsfonts}

\begin{document}


\title{Uniqueness of the Fock representation of the Gowdy
$S^1\times S^2$ and $S^3$ models}
\author{Jer\'onimo Cortez}\email{jacq@fciencias.unam.mx}
\affiliation{Departamento de F\'\i sica,
Facultad de Ciencias, Universidad Nacional Aut\'onoma de M\'exico,
A. Postal 50-542, M\'exico D.F. 04510, Mexico.}
\author{Guillermo A. Mena Marug\'an}\email{mena@iem.cfmac.csic.es}
\affiliation{Instituto de Estructura de la Materia,
CSIC, Serrano 121, 28006 Madrid, Spain.}
\author{Jos\'e M. Velhinho}\email{jvelhi@ubi.pt}
\affiliation{Departamento de F\'{\i}sica, Universidade
da Beira Interior, R. Marqu\^es D'\'Avila e Bolama,
6201-001 Covilh\~a, Portugal.}

\begin{abstract}

After a suitable gauge fixing, the local gravitational
degrees of freedom of the Gowdy $S^1\times S^2$ and
$S^3$ cosmologies are encoded in an axisymmetric field
on the sphere $S^2$. Recently, it has been shown that
a standard field parametrization of these reduced
models admits no Fock quantization with a unitary
dynamics. This lack of unitarity is surpassed by a
convenient redefinition of the field and the choice of
an adequate complex structure. The result is a Fock
quantization where both the dynamics and the
$SO(3)$-symmetries of the field equations are
unitarily implemented. The present work proves that
this Fock representation is in fact unique inasmuch
as, up to equivalence, there exists no other possible
choice of $SO(3)$-invariant complex structure leading
to a unitary implementation of the time evolution.

\vskip 3mm \noindent
\end{abstract}

\maketitle
\newpage
\renewcommand{\thefootnote}{\fnsymbol{footnote}}

\section{Introduction}
\label{int}

In a series of papers \cite{CM,ccm1,ccm2,ccmv,cmv}, a
Fock quantization of the linearly polarized Gowdy
$T^3$ cosmologies \cite{gowdy} has been put forward
\cite{ccm1,ccm2} and shown to be unique under natural
conditions \cite{ccmv,cmv}. With respect to previous
proposals \cite{pierri,torre}, a crucial step was the
redefinition of the scalar field that effectively
parametrizes the local degrees of freedom of the model
\cite{ccm1,ccm2}. This new parametrization allowed the
construction of a Fock quantization with unitary
dynamics, in contrast to the situation found in
\cite{CM,cmv,torre,non-uni} when the seemingly more
natural parametrization adopted in reference
\cite{pierri} is used.

In addition to a unitary dynamics, the quantization
introduced in \cite{ccm1,ccm2} provides a unitary
representation of the group of symmetries of the
(reduced) model, which in the $T^3$ case is actually a
gauge group. This was achieved by means of a complex
structure that is invariant under the action of these
symmetries.$^{\footnotemark[5]}$\footnotetext[5]{Let
us recall that a quantization of the Fock type is
determined by a complex structure on the space of
classical solutions, and that symplectic
transformations which leave the complex structure
invariant are implemented by unitary transformations
which leave the vacuum invariant (up to a phase), see
e.g. \cite{baez,hr}.} Moreover, it was shown that the
conditions of unitary implementation of the dynamics
and invariance of the complex structure completely fix
the quantization, i.e. any two Fock representations
satisfying these conditions are unitarily equivalent
\cite{ccmv}.

More recently, part of the results obtained originally
in the context of the $T^3$ model were extended to the
linearly polarized Gowdy $S^1\times S^2$ and $S^3$
models \cite{BVV1,BVV2}. As shown in \cite{BVV1}, the
local degrees of freedom of these models are
effectively described by an axisymmetric scalar field
on $S^2$ [more precisely in a space-time
$(0,\pi)\times S^2$], obeying the same field equation
in both cases. Starting from this formulation, the
issue of unitary evolution was then discussed,
restricting the considerations to Fock representations
of the scalar field determined by $SO(3)$-invariant
complex structures \cite{BVV2}. Firstly, it was found
that, like in the $T^3$ case, the seemingly natural
field parametrization of these models does not admit a
quantization with unitary dynamics. Secondly, it was
seen that a field redefinition of the type considered
in the $T^3$ case again allows for a unitary
implementation of the dynamics.

The aim of the present work is to show that the
uniqueness theorem presented in \cite{ccmv}, directly
applicable in the Gowdy $T^3$ case as well as in more
general circumstances, is again valid in the Gowdy
$S^1\times S^2$ and $S^3$ cases (when the new field
parametrization is adopted). Specifically, we will
show that, among the set of complex structures
considered in \cite{BVV2}, those that allow a unitary
implementation of the scalar field dynamics define a
unique unitary equivalence class of representations.
Let us stress that restricting attention to complex
structures (or states) that remain invariant under
symmetry groups is a standard practice in quantum
field theory, as a natural way to ensure the unitary
implementation of those groups. This applies both to
the cases of gauge groups, or simply of symmetries
leading to conservation laws.

The paper is organized as follows. In section
\ref{ccmv-quant} we briefly review the quantization of
the $S^1\times S^2$ and $S^3$ models along the lines of
\cite{BVV2}. In section \ref{sec:3} we show the
uniqueness of the quantization. This is the main
section of the paper. The proof of this uniqueness
result is an adaptation of the one presented in
\cite{ccmv}. To avoid unnecessary repetitions, only
the essential technical arguments are explained,
obviating a discussion of the framework that can be
found in \cite{CM,ccm1,ccm2,ccmv,cmv}. We present our
conclusions in section \ref{sec:conc-dis}, together
with a brief discussion of other relevant points.

\section{The quantization of the $S^1\times S^2$
and $S^3$ models}
\label{ccmv-quant}

In this section we briefly review the quantization of
the Gowdy $S^1\times S^2$ and $S^3$ models discussed
in \cite{BVV1,BVV2}.

In the classical theory, once the reduction, gauge
fixing and deparametrization of the models have been
performed, the effective configuration variable for
both the Gowdy $S^1\times S^2$ and $S^3$ linearly
polarized cosmologies is an axisymmetric field on the
sphere $S^2$, which after a mode decomposition in
terms of spherical harmonics can be written as
\begin{equation}
\label{1} \phi(t,s)=\sum_{\ell=0}^{\infty}\left[\,
a_{\ell}y_{\ell}(t)Y_{\ell\, 0}(s)+
a_{\ell}^{*}y^{*}_{\ell}(t)Y^{*}_{\ell\, 0}(s)\,
\right].
\end{equation}
Here, $s\in S^2$, $t\in(0,\pi)$ is the (internal)
time, $Y_{\ell\, 0}$ is the $(\ell, m=0)$ spherical
harmonic and the symbol $*$ denotes complex
conjugation.

The field $\phi$ obeys the equation
\begin{equation}
\label{2} \ddot \phi+\cot{t}\,\dot \phi
-\Delta_{S^2}\phi=0,
\end{equation}
where $\Delta_{S^2}$ denotes the Laplace-Beltrami
operator on $S^2$ and the dot stands for the time
derivative. The field equation (\ref{2}) is invariant
under the group $SO(3)$, acting as rotations on $S^2$.

Given equation (\ref{1}), the dynamics of the
system can be described in terms of the infinite set
of modes $\{y_\ell\}$, which from (\ref{2}) satisfy
the equations of motion:
\begin{equation}
\label{3} \ddot y_\ell+\cot t\,\dot y_\ell+
\ell(\ell+1)y_\ell=0.
\end{equation}
Independent solutions of these equations are, for each
mode, the functions $P_\ell(\cos t)$ and $Q_\ell(\cos
t)$, where $P_\ell$ and $Q_\ell$ denote the first and
second class Legendre functions \cite{abra}.

Endowing the space of solutions with a complex
structure $\bar{J}$ (compatible with the symplectic
form) one can construct a Fock representation of the
field $\phi$. In order to preserve the
$SO(3)$-symmetry in the quantum description, one
restricts the attention to the set of complex
structures which descend from $SO(3)$-invariant ones
under the restriction of axisymmetry (since the field
$\phi$ must also be axisymmetric owing to the Killing
symmetries of the models). The result is a family
$\{\bar{\cal{F}}\}$ of $SO(3)$-invariant Fock
representations. However, time evolution fails to be
implemented as a unitary transformation in each member
of $\{\bar{\cal{F}}\}$ \cite{BVV2}.

In order to arrive at a unitary theory, a time
dependent transformation of the basic field is
performed, namely
\begin{equation}
\label{4} \xi:=\sqrt{\sin t}\,\phi,
\end{equation}
which is analogous to the transformation proposed in
\cite{ccm1,ccm2}. The field $\xi$ can be expanded in
terms of the new modes $z_\ell(t):= \sqrt{\sin
t}\,y_\ell(t)$. Since relation (\ref{4}) is simply a
time dependent scaling, the $SO(3)$-transformations
again define dynamical symmetries of the field $\xi$.

Turning now to the Fock quantizations of the field
$\xi$, these are determined by the possible complex
structures on the space of classical solutions
$\{z_\ell,\ell=0,1,2,\ldots\}$ to the mode equations
\begin{equation}
\label{3bis} \ddot z_\ell+\left[\frac{1}{4}(1+\csc^2
t)+ \ell(\ell+1)\right]z_\ell=0.
\end{equation}
Owing to the commented $SO(3)$-symmetries, we will
restrict our attention to the class of complex
structures which descend from $SO(3)$-invariant ones.
As shown in \cite{BVV2}, this class is parametrized by
sequences of real pairs $\{(\rho_\ell,\nu_\ell)\}$,
where $\rho_\ell>0$ $\forall \ell$. To be precise, let
us consider the complex combinations of classical
solutions
\begin{equation}
\label{n1} z^J_\ell(t)=\left[\rho_\ell P_\ell(\cos
t)+\left(\nu_\ell+\frac{i}{\rho_\ell}\right)
Q_\ell(\cos t)\right] \sqrt{\frac{\sin{t}}{2}}.
\end{equation}
Then, the complex structure $J$ defined by the pairs
$(\rho_\ell,\nu_\ell)$ is such that
\begin{equation}
\label{n2} J\left(z^J_\ell\right)=iz^J_\ell,\ \ J
\big( z^{J\,*}_\ell \big)=-i z^{J\,*}_\ell.
\end{equation}

The unitarity of the dynamics of the field $\xi$
depends on the quantum representation, and therefore
on the complex structure $J$ which determines it. For
each $J$ and for each pair $t_0$, $t_1\in (0,\pi)$,
the symplectic transformation defined by classical
evolution from time $t_0$ to time $t_1$ is determined
by the Bogoliubov coefficients
\cite{BVV2}:$^{\footnotemark[6]}$
\footnotetext[6]{Reference \cite{BVV2} adopts a non
standard notation for the Bogoliubov coefficientes,
which we also follow here to avoid confusions. The
standard coefficients are $-i\alpha^J_\ell$ and
$-i\beta^{J\,*}_\ell$.} \begin{eqnarray}
\label{n3} \alpha_\ell^J(t_0,t_1)&=& z^J_\ell(t_1)
\left[{\dot z}^{J\,*}_\ell(t_0)-\frac{1}{2} \cot t_0\,
z^{J\,*}_\ell(t_0)\right] \nonumber \\ &-&
z^{J\,*}_\ell(t_0)\left[{\dot
z}^J_\ell(t_1)-\frac{1}{2}
\cot t_1\, z^J_\ell(t_1)\right],\nonumber \\
\beta_\ell^J(t_0,t_1)&=& z^J_\ell(t_1) \left[{\dot
z}^J_\ell(t_0)-\frac{1}{2} \cot t_0\,
z^J_\ell(t_0)\right] \nonumber \\ &-&
z^J_\ell(t_0)\left[{\dot z}^J_\ell(t_1)-\frac{1}{2}
\cot t_1\, z^J_\ell(t_1)\right],
\end{eqnarray}
with $\alpha_\ell^J(t_0,t_1)$ and
$\beta_\ell^J(t_0,t_1)$ being the linear and
antilinear part of the transformation, respectively.

It follows from well-known general results
\cite{sh,hr} that the evolution from $t_0$ to $t_1$ is
unitarily implementable in the Fock representation
defined by the complex structure $J$ iff the sequence
$\{\beta_\ell^J(t_0,t_1)\}$ is square summable (SQS);
i.e., the dynamics is unitarily implementable iff
$\sum_{\ell=0}^{\infty}|\beta_\ell^J(t_0,t_1)|^2<\infty$
for all $t_0$, $t_1\in (0,\pi)$.

Employing the asymptotic expansion of the Legendre
functions for large values of
$\ell$,{$^{\footnotemark[7]}$\footnotetext[7]{Note
that the first subdominant terms in these expansions
are  of order $O(\ell^{-3/2})$.}} given e.g. in \cite{GR}, one can see that
this condition of square summability is satisfied for
a large subclass of $SO(3)$-invariant complex
structures, which includes in particular
the complex structure determined by
$\rho_\ell=\sqrt{\pi/2}$ and $\nu_\ell=0$ $\forall
\ell$.

\section{Uniqueness of the quantization}
\label{sec:3}

Let $J_F$ denote the complex structure defined by the
particular values $\rho_\ell=\sqrt{\pi/2}$ and
$\nu_\ell=0$ $\forall \ell$. We will call
$\{z^{J_F}_\ell,z^{J_F\,*}_\ell\}$ the set of complex
classical solutions associated with $J_F$. On the
other hand, let us introduce the following parameters
$A_\ell$ and $B_\ell$:
\begin{eqnarray}
A_\ell &:=& \frac{1}{\sqrt{2\pi}}\left[\rho_\ell-
i{\pi\over 2}\left(\nu_\ell+\frac{i}{\rho_\ell}\right)
\right],\nonumber \\
B_\ell &:=& \frac{1}{\sqrt{2\pi}}\left[\rho_\ell+
i{\pi\over 2}\left(\nu_\ell+\frac{i}{\rho_\ell}\right)
\right]. \label{6}
\end{eqnarray}
These parameters provide the transformation from
$\{z^{J_F}_\ell,z^{J_F\,*}_\ell\}$ to the set of
solutions $\{z^{J}_\ell,z^{J\,*}_\ell\}$ which
corresponds to the complex structure $J$ determined by
$\{(\rho_\ell,\nu_\ell)\}$, namely
\begin{equation}
\label{z1} z^J_\ell=A_\ell\, z^{J_F}_\ell + B_\ell\,
z^{J_F\,*}_\ell.
\end{equation}
Note also that
\begin{equation}
\label{9} |A_\ell|^2-|B_\ell|^2=1\ \ \forall \ell.
\end{equation}
It then follows that $|A_\ell|\geq 1$ $\forall \ell$,
and that the sequence $\{B_\ell/A_\ell\}$ is bounded.

More importantly, as a consequence of transformation
(\ref{z1}), one concludes that the complex structures
$J$ [with parameters $\{(A_\ell , B_\ell)\}$] and
$J_F$ determine unitarily equivalent Fock
representations iff the sequence $\{B_\ell\}$ is SQS
(see e.g. \cite{ccmv} for details).

We will now prove that if a complex structure $J$ is
such that the sequence $\{\beta^J_\ell(t_0,t_1)\}$ is
SQS $\forall t_0$, $t_1$, then the sequence
$\{B_\ell\}$ is necessarily SQS, so that the
representations determined by $J$ and $J_F$ are
equivalent. In other words, we will prove that the
Fock representation selected by $J_F$ is the unique
(up to unitary equivalence) $SO(3)$-invariant Fock
representation where the dynamics is implemented as a
unitary transformation.

In order to simplify the notation, the sequences
$\{\beta_\ell^{J_F}(t_0,t_1)\}$ and
$\{\alpha_\ell^{J_F}(t_0,t_1)\}$ will be respectively
denoted from now on $\{\beta_\ell(t_0,t_1)\}$ and
$\{\alpha_\ell(t_0,t_1)\}$. It is not difficult to see
that the coefficients $\beta^J_\ell$, $\alpha_\ell$
and $\beta_\ell$ are related by
\begin{equation}
\beta^J_\ell(t_0,t_1)=A^2_\ell\beta_\ell(t_0,t_1)+
B^2_\ell\beta^{*}_\ell(t_0,t_1)+ 2A_\ell B_\ell\,
{\rm{Re}}[\alpha_\ell(t_0,t_1)]. \label{8}
\end{equation}
Here, ${\rm{Re}}[\, .\, ]$ denotes the real part. Let
us then suppose that $\{\beta^J_\ell(t_0,t_1)\}$ is
SQS $\forall t_0$, $t_1\in (0,\pi)$, so that the
dynamics is unitarily implemented in the Fock
representation determined by the $SO(3)$-invariant
complex structure $J$. Then, since $|A_\ell|\geq 1$,
the sequence $\{ \beta^J_\ell(t_0,t_1)/ A^2_\ell\}$ is
also SQS. We have
\begin{equation}
{\beta^J_\ell(t_0,t_1)\over A^2_\ell} =
\beta_\ell(t_0,t_1)+{B^2_\ell\over A^2_\ell}
\beta^{*}_\ell(t_0,t_1)+2 {B_\ell\over A_\ell}\,
{\rm{Re}}[\alpha_\ell(t_0,t_1)]. \label{10}
\end{equation}
Given that $\{\beta_\ell(t_0,t_1)\}$ is SQS and the
sequence $\{B^2_\ell/A^2_\ell\}$ is bounded, it
follows that $\{\beta_\ell(t_0,t_1)+(B^2_\ell/
A^2_\ell) \beta^{*}_\ell(t_0,t_1)\}$ is SQS. Hence,
since the space of SQS sequences is a linear space,
one concludes that the sequence $\{(B_\ell/ A_\ell)\,
{\rm{Re}}[\alpha_\ell(t_0,t_1)]\}$ is SQS $\forall
t_0$, $t_1\in(0,\pi)$.

Using the asymptotic expansion of the Legendre
functions for large $\ell$ \cite{GR}, one can check that the difference
between ${\rm{Re}}[\alpha_\ell(t_0,t_1)]$ and
$\sin[(\ell+1/2)(t_1-t_0)]$ is a SQS sequence $\forall
t_0$, $t_1\in[\epsilon,\pi-\epsilon]$, where
$\epsilon>0$ is arbitrarily small. Thus, from the
bounds on $B_\ell/A_\ell$ and linearity, one gets that
$\{(B_\ell/A_\ell) \sin[(\ell+1/2)(t_1-t_0)]\}$ is
also SQS. Introducing the notation $T:=t_1-t_0$, one
then concludes that the limit
\begin{equation}
\label{11}
\lim_{N\to\infty}\sum_{\ell=0}^{N}\frac{|B_\ell|^2}
{|A_\ell|^2}
\sin^2\left[\left(\ell+\frac{1}{2}\right)T\right]=:f(T)
\end{equation}
exists $\forall T\in[0,\pi-\varepsilon]$, with
$\varepsilon:= 2\epsilon$ an arbitrarily small
positive number.

One can now apply the Luzin theorem \cite{luzin},
which ensures that, for every $\delta>0$, there exist
a measurable set $E_{\delta}\subset
[0,\pi-\varepsilon]$ with
$\int_{\overline{E_{\delta}}} dT<\delta$ and a
function $\phi_{\delta}(T)$, continuous on
$[0,\pi-\varepsilon]$, which coincides with $f(T)$ on
$E_{\delta}$. Here, $\overline{E_{\delta}}$ denotes
the complement set $[0,\pi-\varepsilon]\backslash
E_{\delta}$. One then gets
\begin{equation}
\label{12} \sum_{\ell=0}^{N}{|B_\ell|^2 \over
|A_\ell|^2}\int_{E_{\delta}}
\sin^2\left[\left(\ell+\frac{1}{2}\right)T\right]dT\leq
\int_{E_{\delta}}f(T)dT=:I_{\delta}\quad \forall N,
\end{equation}
where $I_{\delta}=\int_{E_{\delta}}\phi_{\delta}(T)dT$
is some finite number, and the inequality follows from
the fact that $f(T)$ is the limit of an increasing
sequence, given by sums of nonnegative terms. On the
other hand, one finds \begin{eqnarray}
\label{13}\int_{E_{\delta}}
\sin^2\left[\left(\ell+\frac{1}{2}\right)T\right]dT &
= &  \int_0^{\pi}
\sin^2\left[\left(\ell+\frac{1}{2}\right)T\right]dT -
\int_{\pi-\varepsilon}^{\pi}
\sin^2\left[\left(\ell+\frac{1}{2}\right)T\right]dT
\nonumber \\
&-& \int_{\overline{E_{\delta}}}
\sin^2\left[\left(\ell+\frac{1}{2}\right)T\right]dT
\geq {\pi\over 2}-\varepsilon-\delta \quad\quad
\forall \ell.
\end{eqnarray}
Combining (\ref{12}) and (\ref{13}) one obtains
\begin{equation}
\label{14} I_{\delta}\geq \sum_{\ell=0}^{N}{|B_\ell|^2
\over |A_\ell|^2} \left({\pi\over
2}-\varepsilon-\delta\right)\quad \forall N.
\end{equation}
Since it is clearly possible to choose $\delta$ and
$\varepsilon$ such that $\pi-2\varepsilon-2\delta>0$,
one concludes that
\begin{equation}
\label{15} \sum_{\ell=0}^{N}{|B_\ell|^2 \over
|A_\ell|^2} \leq {2 I_{\delta}\over
\pi-2\delta-2\varepsilon} \quad \forall N,
\end{equation} implying that the infinite sum
$\sum_{\ell=0}^{\infty}(|B_\ell|^2/|A_\ell|^2)$
exists.

Finally, since $\{B_\ell/A_\ell\}$ is SQS, the ratio
$B_\ell/A_\ell$ necessarily tends to zero. In
particular, it then follows from (\ref{9}) that the
sequence $\{A_\ell\}$ is bounded. Therefore, the
sequence $\{B_\ell=A_\ell (B_\ell/A_\ell)\}$ is also
SQS, as we wanted to prove.

\section{Conclusion and further comments}
\label{sec:conc-dis}

The discussion presented in this work provides a
natural extension to the $S^1 \times S^2$ and $S^3$
topologies of the uniqueness result obtained in
\cite{ccmv} for the Fock quantization of the Gowdy
$T^3$ model. For those other topologies we have proved
that, among the set of complex structures that are
invariant under the group of $SO(3)$-symmetries of the
reduced model, there exists a unique unitary
equivalence class such that the field evolution is
implemented in the quantum theory as a unitary
transformation. We have selected as representative for
this unique class the complex structure $J_F$ defined
by the particular values $\rho_\ell=\sqrt{\pi/2}$ and
$\nu_\ell=0$ $\forall \ell$. The associated set of solutions
$\{ z^{J_F}_\ell, z^{J_F}_\ell\,^* \}$ can be obtained from
(\ref{n1}). It is easy to see that, for large $\ell$,
these solutions have the asymptotic behavior
\begin{equation}
z^{J_F}_\ell = \frac{1}{\sqrt{2\ell+1}} e^{-i \left(\ell +
\frac{1}{2}\right)t+i\frac{\pi}{4}}+ O(\ell^{-3/2}).
\end{equation}
Disregarding the subdominant correction $O(\ell^{-3/2})$,
these are precisely the solutions that one would
obtain for equation (\ref{3bis}) in the case that the
time dependent potential term (proportional to $\csc^2
t$) could be neglected, situation that would
correspond to a stationary field equation for $\xi$.

The uniqueness proof given here is an extension of our
proof for the Gowdy $T^3$ model explained in
\cite{ccmv}. Apart from adapting some steps of the
demonstration to deal with other topologies, the
present proof differs from the previous version in a
partial simplification of the arguments, achieved
mainly by realizing that the subdominant terms in
${\rm Re}[\alpha_\ell(t_0,t_1)]$ for large $\ell$
provide in fact a SQS sequence.

Another issue that we would like to comment is the
freedom in the choice of momentum conjugate to the
scalar field $\xi$. The choice made in \cite{BVV2} has
the problem of leading to a Hamiltonian that contains
a contribution that is linear in the momentum. It is
seen in \cite{BVV2} that, in the Fock quantization
defined by $J_F$, the natural vacuum of the theory
does not belong to the domain of the normal ordered
Hamiltonian. Nonetheless, one can introduce a change
of momentum of the form $P_{(N)}= P + \cot{t} \;Q/2$,
where $Q$ and $P$ are the field and its momentum
evaluated in the section of constant time under
consideration. This change can be understood as a time
dependent canonical transformation. It leads to a new
Hamiltonian $H_0$ that is quadratic both in $Q$ and in
$P_{(N)}$ and such that its action on the vacuum is
well defined. In fact, this time dependent change of
momentum can be alternatively understood as the result
of a canonical transformation performed before the
deparametrization of the model. The reduced
Hamiltonian that one obtains from the Hilbert-Einstein
action by means of that canonical transformation and
the subsequent deparametrization is precisely the
Hamiltonian $H_0$ alluded above.

Finally, the fact that the vacuum of the Fock
representation is contained in the domain of the
reduced Hamiltonian may be of practical importance for
the success of certain quantization approaches. This
affects not only the possibility of defining the
action of the evolution operator on the vacuum (in the
Schr\"odinger picture) as a formal series in powers of
the Hamiltonian, but may also be relevant in
quantization schemes that introduce discretizations in
which the evolution operator is bound to be
substituted by a repeated action of the Hamiltonian,
as it may happen to be the case in Loop Quantum
Cosmology \cite{MMV}.

\bigskip
\bigskip

\baselineskip = 17 pt
{\em  Note added in proof.} {  During the consideration of this work for publication, the authors came to know that a discussion about the uniqueness of the quantization was included in the final version of [13]. That treatment is, however, incomplete and not entirely correct. 
On  one hand, only a small subclass of the set of complex structures that allow unitary dynamics is considered. On the other hand, the first condition in equation (4.22) is in fact not sufficient for a unitary dynamics or for the unitary equivalence of the representations.} 

\baselineskip = 19 pt
\section*{Acknowledgements}
This work was supported by the Spanish MEC Project
FIS2005-05736-C03-02, the Spanish Consolider-Ingenio
2010 Programme CPAN (CSD2007-00042), the Joint
CSIC/CONACyT Project 2005MX0022 and the Portuguese FCT
Project POCTI/FIS/57547/2004.

\end{document}